\documentclass[sigconf]{acmart-me}

\usepackage{booktabs} 
\usepackage{url}
\usepackage{color}
\usepackage{enumitem}
\usepackage{multirow}
\setcopyright{rightsretained}

\acmDOI{}

\acmISBN{}

\acmConference[MediaEval'19]{Multimedia Evaluation Workshop}{27-29 October 2019}{Sophia Antipolis, France} 
\acmYear{}
\copyrightyear{}
\acmPrice{}

\begin{document}
\title{Music theme recognition using CNN and self-attention}

\author{Manoj Sukhavasi, Sainath Adapa}
\email{manoj.sukhavasi1@gmail.com, adapasainath@gmail.com}

\renewcommand{\shortauthors}{M. Sukhavasi, S. Adapa}
\renewcommand{\shorttitle}{Emotion and Theme recognition in music using Jamendo}

\begin{abstract}
We present an efficient architecture to detect mood/themes in music tracks on \texttt{autotagging-moodtheme} subset of the MTG-Jamendo dataset. Our approach consists of two blocks, a CNN block based on MobileNetV2 architecture and a self-attention block from Transformer architecture to capture long term temporal characteristics. We show that our proposed model produces a significant improvement over the baseline model. Our model (team name: AMLAG) achieves 4\textsuperscript{th} place  on PR-AUC-macro Leaderboard in \texttt{MediaEval 2019: Emotion and Theme Recognition in Music Using Jamendo}.
\end{abstract}

\maketitle

\section{Introduction}
\label{sec:intro}

Automatic music tagging is a multi-label classification task to predict the music tags corresponding to the audio contents. Tagging music with themes (action, documentary) and mood (sad, upbeat) can be useful in music discovery and recommendation.
\texttt{MediaEval 2019: Emotion and Theme Recognition in Music Using Jamendo} aims to improve the machine learning algorithms to automatically recognize the emotions and themes conveyed in a music recording \cite{jamendo2019}. This task involves the prediction of moods and themes conveyed by a music track, given the raw audio on the \texttt{autotagging-moodtheme} subset of the MTG-Jamendo dataset \cite{jamendodataset}. The overview paper \cite{jamendo2019} describes the task in more detail, and also introduces us to a baseline solution based on VGG-ish features. In this paper, we describe our Fourth place submission on PR-AUC-macro Leaderboard \footnote{\url{https://multimediaeval.github.io/2019-Emotion-and-Theme-Recognition-in-Music-Task/results}} which improves the results significantly on the baseline solution.

\section{Related Work}
\label{sec:work}

Conventionally feature extraction from audio relied on signal processing to compute relevant features from time or frequency domain representation. As an alternative to these solutions, architectures based on Convolutional Neural Networks(CNN)  \cite{choi2016} have become more popular recently following their success in CV, speech processing. Extensions to CNNs have also been proposed to capture the long term temporal information in the form of CRNN \cite{choi2016crnn}. Recently \cite{won2019selfatt} has shown that self-attention applied to music tagging captures temporal information. This architecture was based on the transformer architecture which was very successful in Natural Language Processing (NLP)\cite{vaswani2017transformer}. In this paper, we propose two methods MobileNetV2 and MobileNetV2 with self-attention which are based mainly on these two previous works \cite{adapa2019urban, won2019selfatt}.

\section{Approach}
\label{arch}

We used the pre-computed Mel-spectrograms made available by the organizers of the challenge\footnote{https://github.com/MTG/mtg-jamendo-dataset}. No additional pre-processing steps were undertaken other than the normalization of the input Mel-spectrogram features.

As image-based data augmentation techniques have been shown to be effective in audio tagging \cite{adapa2019urban, freesound2019position1}, we used transformations such as Random crop and Random Scale. Additionally, we also employed SpecAugment and Mixup. SpecAugment\cite{park2019specaugment} proposed initially for speech recognition,  masks blocks of frequency channels or time steps of a log Mel-spectrogram. Mixup \cite{zhang2017mixup} samples two training examples randomly and linearly mixes them (both the feature space and the labels).

We propose two methods: MobilenetV2 architecture, and MobileNetV2 architecture combined with a self-attention block to capture long term temporal characteristics. We describe both of these methods in detail below.

\subsection{MobileNetV2}
\label{arch1}

It has been shown previously that using pre-trained ImageNet models helps in the case of audio tagging \cite{adapa2019urban, lasseck2018acoustic}. Hence, we employed MobileNetV2 \cite{sandler2018mobilenetv2} for the current task. Since Mel-spectrograms are single channel, the input data is transformed into a three-channel tensor by passing it through two convolution layers. This tensor is then sent to the MobileNetV2 unit. As the number of labels is different here, the linear layer at the very end is replaced. No other modifications were performed to the original MobileNetV2 architecture.

\subsection{MobileNetV2 with Self-attention}
\label{arch2}
 
 The architecture described in sub-section \ref{arch1} might not be able to capture the long-term temporal characteristics. The dataset consists of tracks with varying lengths with a majority longer than 200s. Self-attention has been shown to capture long-range temporal characteristics in the context of music tagging \cite{won2019selfatt}. Hence self-attention mechanism can be helpful in the current task. In this section, we describe our extended MobileNetV2 architecture with self-attention.
 
 The architecture consists of 2 main blocks: modified MobileNetV2 (identical to the architecture described in \cite{adapa2019urban}) to capture freq-time characteristics, and the self-attention block to capture long term temporal characteristics.
 
 Similar to the transformer model \cite{vaswani2017transformer}, multi-head self-attention with positional encoding was implemented for the current architecture. Since our task consists only of classification we use only the encoder part of it similar to BERT \cite{dev2018bert}. Our implementation is based on the architecture described in \cite{won2019selfatt}. We use 4 attention heads and 2 attention layers. The input sequence length is 16 and has embedding size of 256.

The control flow within this architecture is as follows:
\begin{itemize}
\item Input is a Mel-spectrogram tensor of length 4096 (number of bands being 96). This input is divided length-wise into 16 segments, with each segment's length being 256.
\item Each of the 16 slices is sent through the modified MobileNetV2 block to extract the features.
\item The feature maps are then fed into the Self-attention block. At the end of this block, two dense layers are put to use to generate the predictions.
\item Additionally, the feature maps from the MobileNetV2 block are also used to generate predictions. With each segment, we have a set of predictions. All the sixteen predictions are averaged to obtain the final prediction.
\end{itemize}

As described above, the architecture generates two predictions: one solely using the MobileNetV2, and the other using the MobileNetV2 and the Self-attention blocks. While training, combined loss from both the predictions are used for back-propagation.

\section{Training and results}
\label{results}

We made two submissions under the team name AMLAG\footnote{\url{https://github.com/sainathadapa/mediaeval-2019-moodtheme-detection}}, one each using the two architectures described in sections \ref{arch1} and \ref{arch2}. Both the submissions employ the same Mel-spectrogram inputs and Binary Cross-entropy loss as the optimization metric. PyTorch \cite{paszke2017automatic} was used for training the model in both cases.

\begin{figure}[]
\raggedleft
\includegraphics[scale=0.85]{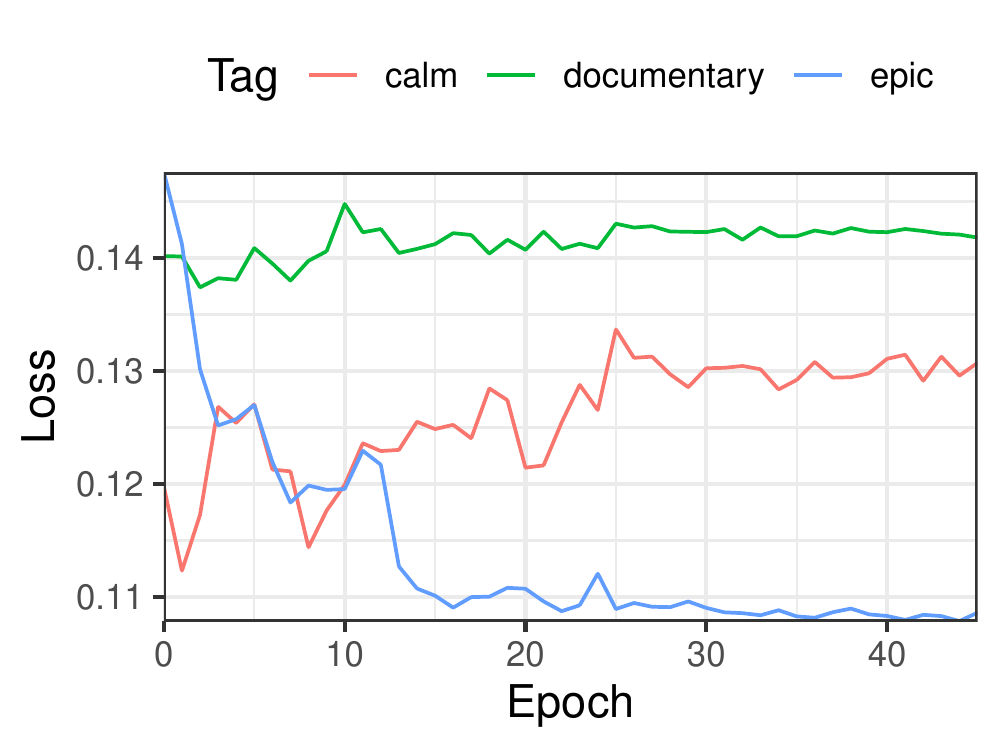}
\caption{Trend in loss values for three sample classes while training the MobileNetV2 model. The plot illustrates the detail that not all classes were benefiting from joint training. In this case, the loss for \textit{epic} class is decreasing while the loss for \textit{calm} is increasing, \textit{documentary} loss is almost stagnant.}
\label{fig:loss_plot}
\end{figure}

For \textit{submission 1}, the AMSGrad variant of the Adam algorithm \cite{kingma2014adam, reddi2019convergence} with a learning rate of 1e-3 was utilized for optimization. Whenever the overall loss on the validation set stopped improving for five epochs, the learning rate was reduced by a factor of 10. For this training we use input Mel-spectrogram of length 6590, padding is used to make all the inputs of constant length. We observed that not all classes benefited from being trained together (see Figure \ref{fig:loss_plot}). Hence, following the approach taken in \cite{caruana1998dozen}, early stopping was done separately for each class based on the loss value for that particular class. Additionally, an attempt was made to find subsets of classes that train well together, but ultimately the overall performance had been lower than when all the classes were jointly trained. This is one avenue for future research with this dataset.

To prepare \textit{submission 2},  we use input Mel-spectrogram of length 4096, padding is used to make all the inputs of constant length. We train the model for 120 epochs while utilizing Adam as the initial optimizer. We then employ an optimization technique proposed in \cite{nit2017adamsgd, won2019selfatt}: the optimizer is switched from Adam to Stochastic gradient descent (with Nesterov momentum \cite{sutskever2013importance}) after 60 epochs for better generalization of the model. Early stopping was done jointly for all classes based on the macro-averaged AUC-ROC on the validation set.

We present the results for both the submissions in Table \ref{fig:results}. Also, results from the baseline approach that uses VGG-ish architecture are shown for comparison purposes. In all the metrics, the MobileNetV2 with a self-attention block exhibits an improvement over solely using the MobileNetV2. With respect to the baseline model, \textit{submission 2} proved to be an improvement over all but the micro-averaged F-score and Precision metrics. On the task leaderboard, our model achieved 4\textsuperscript{th} position in case of PR-AUC-macro, and 5\textsuperscript{th} position in case of F-score-macro.

\begin{table}[]
\centering
\begin{tabular}{llll}
                & \begin{tabular}[c]{@{}l@{}}\textbf{Baseline}\\ \textbf{(vggish)}\end{tabular} & \begin{tabular}[c]{@{}l@{}}\textbf{Submi-}\\ \textbf{ssion 1}\end{tabular} & \begin{tabular}[c]{@{}l@{}}\textbf{Submi-}\\ \textbf{ssion 2}\end{tabular} \\
\toprule
PR-AUC-macro    & 0.107734          & 0.118306    & 0.125896    \\
ROC-AUC-macro   & 0.725821          & 0.732416    & 0.752886    \\
F-score-macro   & 0.165694          & 0.151891    & 0.182957    \\
precision-macro & 0.138216          & 0.135673    & 0.145545    \\
recall-macro    & 0.30865           & 0.306015    & 0.39164     \\
PR-AUC-micro    & 0.140913          & 0.150605    & 0.151706    \\
ROC-AUC-micro   & 0.775029          & 0.784128    & 0.797624    \\
F-score-micro   & 0.177133          & 0.152349    & 0.164375    \\
precision-micro & 0.116097          & 0.098133    & 0.10135     \\
recall-micro    & 0.37348           & 0.340428    & 0.434691   \\
\toprule
\end{tabular}
\caption{Performance on the test dataset}
\label{fig:results}
\end{table}

\section{Other approaches}

Some of the approaches that we have tried, but haven't observed better performance are listed below:
\begin{itemize}
\item A dense layer architecture that uses OpenL3 embeddings \cite{cramer2019look}
\item A dense layer architecture that uses the pre-computed statistical features from Essentia using the feature extractor for AcousticBrainz. This data was made available by the organizers, along with the raw audio and Mel-spectrogram data.
\item CNN architecture that directly uses the raw audio representation, as described in \cite{kim2019comparison}
\item Similar to using the MobileNetV2 in Section \ref{arch1}, we tested another ImageNet pre-trained architecture - ResNeXt model \cite{xie2017aggregated}.
\end{itemize}

\bibliographystyle{ACM-Reference-Format}
\def\bibfont{\small} 
\bibliography{sigproc} 

\end{document}